\documentclass[12pt]{iopart}
\usepackage{graphicx}
\begin{document}

\title{The properties of the D-meson in dense matter}

\author{Laura Tol\'os$^1$, J\"urgen Schaffner-Bielich$^1$ and Amruta Mishra$^2$}

\address{$^1$Institut
f\"ur Theoretische Physik, J.W. Goethe-Universit\"at $\&$  FIAS \\
Postfach 11 19 32. 60054 Frankfurt am Main, Germany}
\address{$^2$Department of Physics, I.I.T. Delhi, New Delhi - 110 016, India}

\begin{abstract}
We study the D-meson spectral density in dense matter within the framework of a coupled-channel self-consistent calculation taking, as bare meson-baryon interaction, a separable potential. Our coupled-channel model generates dynamically the $\Lambda_c(2593)$ resonance. The medium modifications of the D-meson properties due to Pauli blocking and the dressing of D-mesons, nucleons and pions are also discussed. We found that the coupled-channel effects in the  self-consistent process  reduce the in-medium effects on the D-meson compared to previous works.

\end{abstract}
\vspace{-0.8cm}




\section{Introduction}
The study of medium  modifications of the D-meson has become a subject of recent interest because the important consequences for open-charm enhancement in nucleus-nucleus collisions \cite{cassing} as well as for $J/\Psi$ suppression \cite{NA501}.
The NA50 Collaboration \cite{NA50e} has observed an enhancement of dimuons in Pb+Pb collisions 
which has been attributed to an open-charm enhancement in nucleus-nucleus collisions relative to proton-nucleus reactions. On the other hand, an appreciable contribution for the $J/\Psi$ suppression is expected to be due to the formation of the quark-gluon plasma \cite{blaiz}. However, the suppression could also be understood in an hadronic environment due to  inelastic comover scattering  and, therefore, the medium modification of the D-mesons should modify the $J/\Psi$ absorption \cite{blaschke}. 
 Finally, the D-mesic nuclei, predicted by the quark-meson coupling (QMC) model \cite{qmc}, could also give us information about the in-medium properties of the D-meson. It is shown that the $D^-$ meson forms narrow bound states with $^{208} Pb$ while the $D^0$ is deeply bound in nuclei. It is thus of importance to understand the interactions of the D-mesons in the hadronic medium.

Calculations based on the QCD sum-rule (QSR) approach \cite{arata} as well as on the QMC model \cite{qmc} predict the mass drop of the D-meson to be of the order of 50-60 MeV at nuclear matter density. A similar drop at finite temperature is suggested from the lattice calculations for heavy-quark potentials \cite{digal} together with a recent work based on a chiral model \cite{amruta}. In this paper, we study the spectral density of a D-meson embedded in dense matter incorporating the coupled-channel effects as well as the dressing of the intermediate propagators which have been ignored in the previous works. These effects will turn out to be crucial for describing the D-meson in dense matter \cite{tolos04}.

\section{Formalism}
In order to obtain the D-meson self-energy in dense matter and, hence, the spectral density, the knowledge of the in-medium DN interaction is required. This amplitude is obtained taking, as a bare interaction, a separable potential model
\begin{eqnarray}
V_{i,j}(k,k')=\frac{g^2}{\Lambda^2}C_{i,j}\Theta(\Lambda-k)\Theta(\Lambda-k') \ ,  
\end{eqnarray}
where $g$ and  $\Lambda$ are the coupling constant and cutoff, respectively. These
two parameters will be determined by fixing the position and the width of the $\Lambda_c(2593)$ resonance. For the interaction matrix $C_{ij}$, we use the result derived from SU(3) flavor symmetry
. We are, therefore, confronted with a coupled-channel problem since this interaction allows for the transition from DN to other channels, namely, $\pi \Lambda_c$, $\pi \Sigma_c$, $\eta \Lambda_c$ and $\eta \Sigma_c$, all having charm $c=1$.
With this structure in mind, the G-matrix is given by
\begin{eqnarray}
&&\langle M_1 B_1 \mid G(\Omega) \mid M_2 B_2 \rangle = \langle M_1 B_1
\mid V \mid M_2 B_2 \rangle + \nonumber \\
&&\hspace{-1cm}\sum_{M_3 B_3} \langle M_1 B_1 \mid V \mid
M_3 B_3 \rangle
\frac {Q_{M_3 B_3}}{\Omega-E_{M_3} -E_{B_3}+i\eta} \langle M_3 B_3 \mid
G(\Omega)
\mid M_2 B_2 \rangle \ ,
   \label{eq:gmat1}
\end{eqnarray}
where $M_i$ and $B_i$  represent the possible
mesons (D, $\pi$, $\eta$) and
baryons ($N$, $\Lambda_c$, $\Sigma_c$) respectively, and their corresponding
quantum numbers, and $\Omega$ is the starting energy. The function $Q_{M_3 B_3}$ stands for the Pauli operator while $E_{M_3 (B_3)}$ is the meson (baryon) single-particle energy. 
Then, the D-meson single-particle potential in the Brueckner-Hartree-Fock approach reads
\begin{equation}
 U_{D}(k,E_{D}^{qp})= \sum_{N \leq F} \langle D N \mid
 G_{D N\rightarrow
D N} (\Omega = E^{qp}_N+E^{qp}_{D}) \mid D N \rangle,
\label{eq:self}
\end{equation}
where the summation over nucleonic states is limited by the nucleon Fermi momentum. As it can be seen from Eq.~(\ref{eq:self}), since the DN  interaction ($G$-matrix) depends on the
D-meson single-particle energy, which in turn depends on the
D-meson potential, we are confronted
with a self-consistent problem. After self-consistency for the on-shell value
$U_{D}(k_{D},E_{D}^{qp})$ is
achieved, one can obtain the full self-energy $\Pi_D(k_D,\omega)=2\sqrt{k_D^2+m_D^2} \, U_{D}(k_D,\omega)$. 
 This self-energy can then be used to
determine the
 D-meson single-particle propagator 
 and the corresponding spectral density. 

\begin{figure}[htb]
\begin{minipage}
[t]
  {65mm} 
\includegraphics[width=6cm,height=7cm]{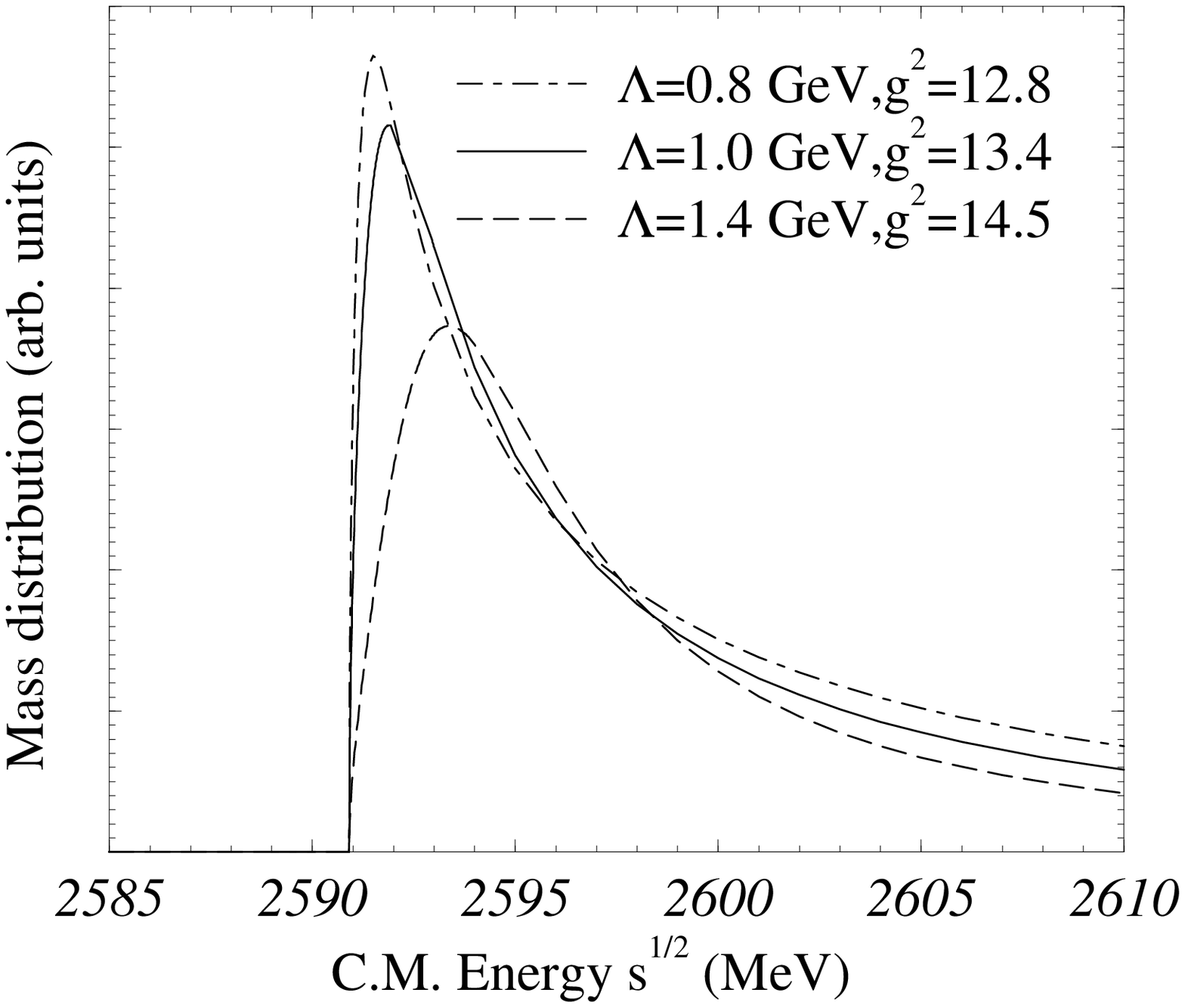}
\end{minipage}
%
%
\begin{minipage}[t]{65mm}
  \includegraphics[width=7cm, height=7cm]{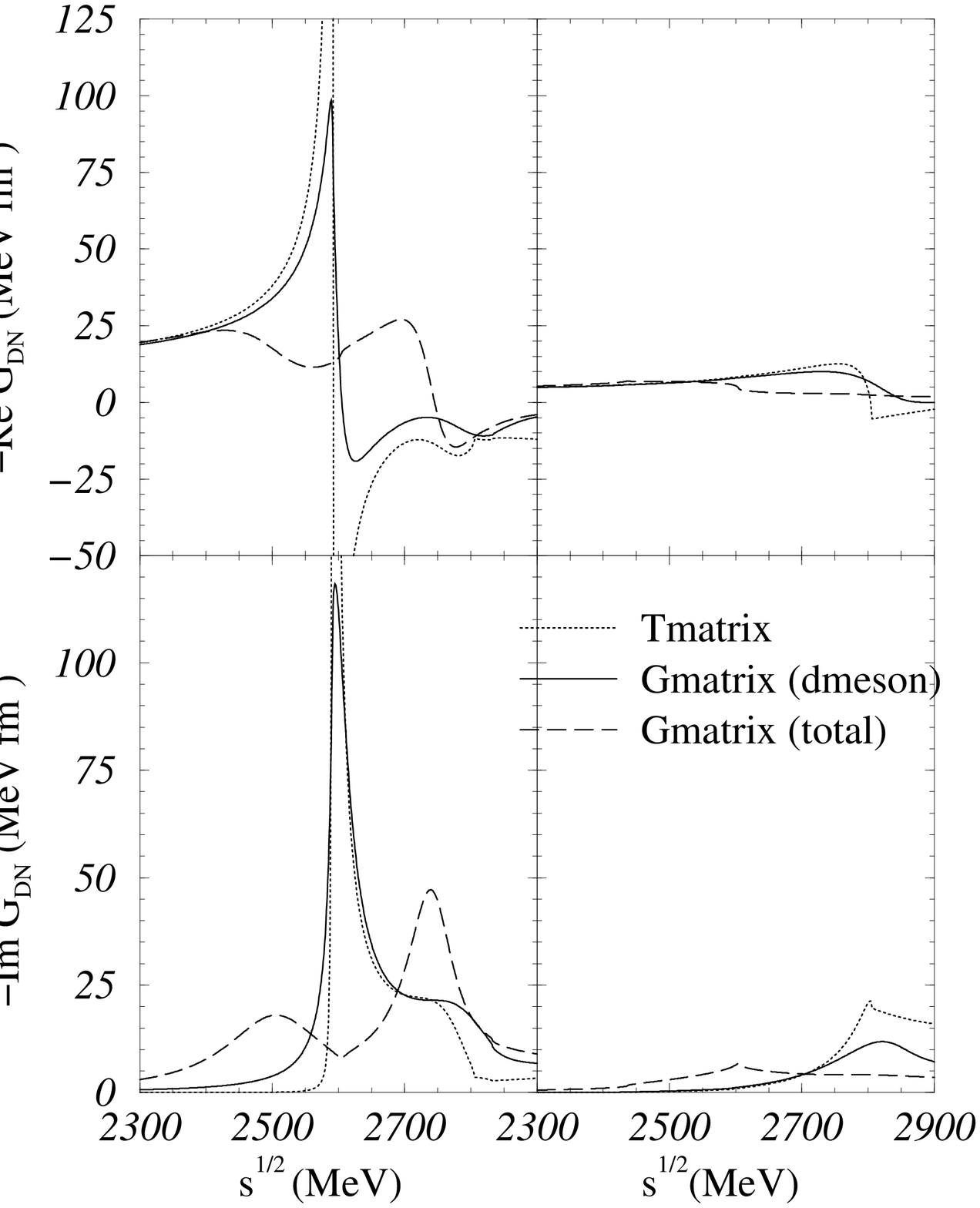}
\end{minipage}
\caption{Left: $\Lambda_c(2593)$ mass spectrum for different sets of coupling constants and cutoffs.
 Right: Real and imaginary parts of the s-wave DN amplitude in the $I=0$ (left panels) and $I=1$ (right panels)  channels as functions of the center-of-mass energy at total momentum zero for $\Lambda=1$ GeV and $g^2=$13.4 and for different approaches: T-matrix calculation (dotted lines), self-consistent calculation for the D-meson at $\rho=\rho_0$  (solid lines) and self-consistent calculation for the D-meson including the dressing of nucleons and the pion self-energy at $\rho=\rho_0$ (long-dashed lines).}
\label{fig:dmeson1}
\end{figure}

\section{Results and Conclusions}

We start this section by showing in the l.h.s. of Fig.~\ref{fig:dmeson1}  the mass distribution  of the $\pi \Sigma_c$ state as a function of the C.M. energy for different sets of  coupling constants $g$ and cutoffs $\Lambda$ (see the definition in Ref.~\cite{tolos04}). 
 Our coupled-channel calculation generates dynamically the $\Lambda_c(2593)$ resonance. The position ($2593.9 \pm 2$ MeV) and width ($\Gamma=3.6^{+2.0}_{-1.3}$ MeV) are obtained for a given set of coupling constants and cutoffs.

Once the position and width of the $\Lambda_c(2593)$ resonance are reproduced dynamically, we study the effect of the different medium modifications on the resonance. In the r.h.s. of Fig.~\ref{fig:dmeson1} we display  the real and imaginary parts of the s-wave DN amplitude for $I=0$ and $I=1$ within 
different approaches.
When the nucleons and pions are dressed, the picture depicted is completely different to the case when only D-mesons are dressed self-consistently.
In this case, for $I=0$,
we observe two structures: one structure  around 2.5 GeV below the $\pi \Sigma_c$ threshold and a second one at 2.8 GeV, which lies below the DN threshold. Both structures are states with the $\Lambda_c$-like quantum numbers.
Whether the first resonant structure  is the in-medium $\Lambda_c(2593)$ resonance 
and the second bump is a new resonance is something that deserves further studies.
\begin{figure}[htb]
\begin{minipage}
[t]
  {65mm} 
\includegraphics[width=7cm,height=6.5cm,angle=-90]{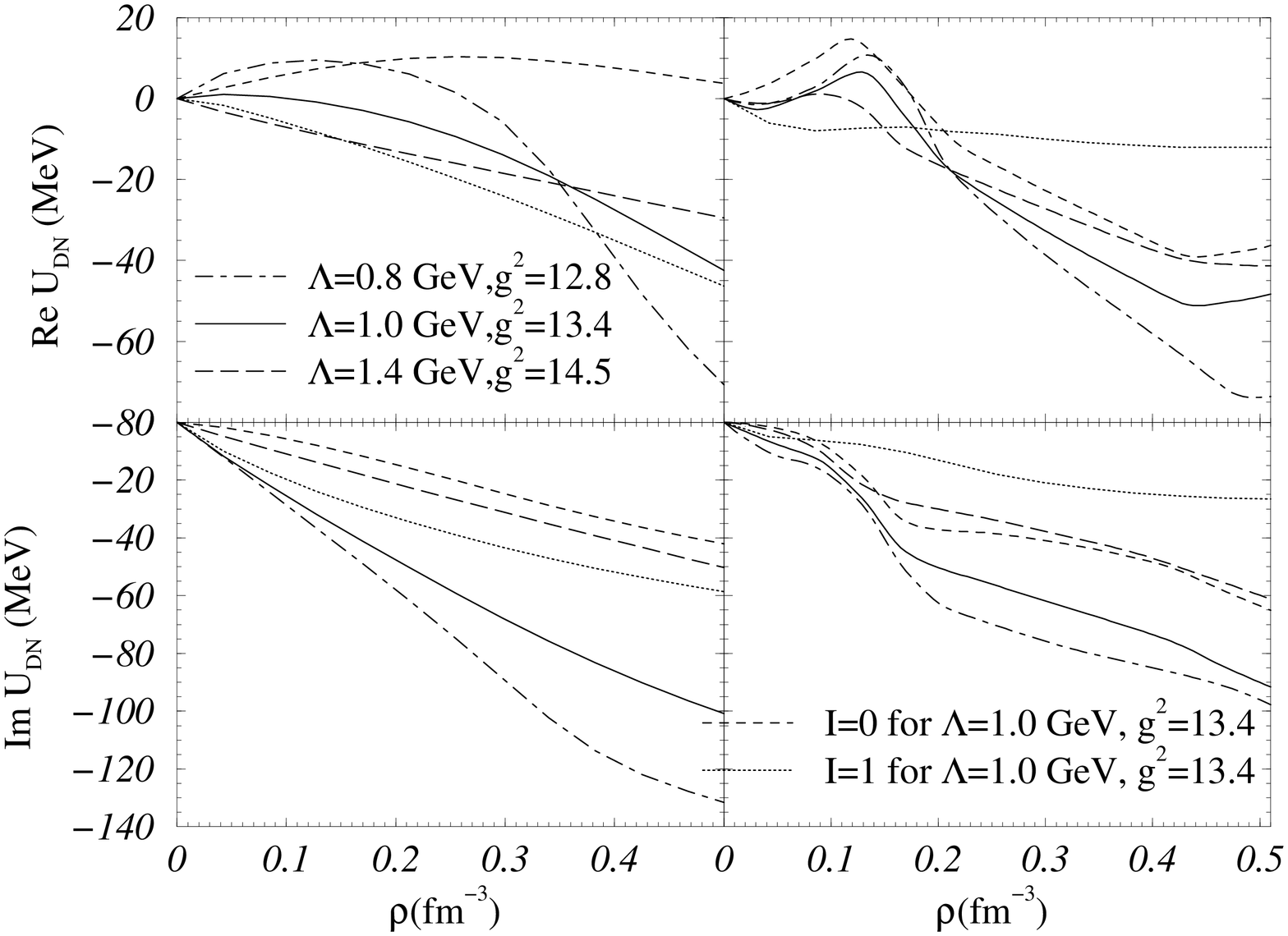}
\end{minipage}
\hspace{\fill}
\begin{minipage}[t]{70mm}
  \includegraphics[width=7cm, height=7cm,angle=-90]{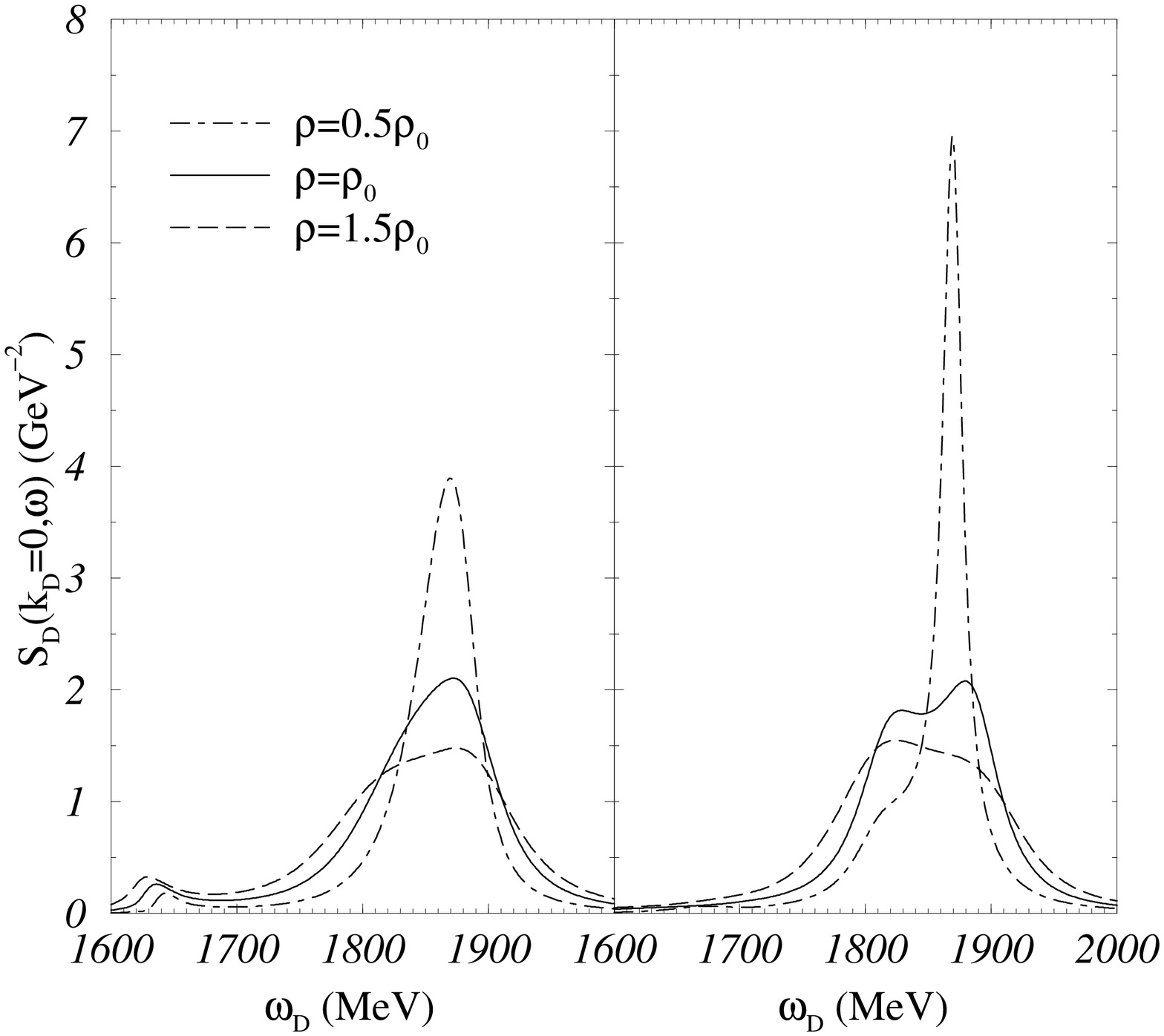}
\end{minipage}
\caption{Left: D-meson  potential at $k_D=0$ as a function of the density, including the isospin decomposition, for different sets of coupling constants and cutoffs and the two self-consistent approaches. Right: 
D-meson spectral density at $k_D=0$ as a function of energy with  $\Lambda=1$ GeV and $g^2=$13.4 for different densities and for the two approaches.}
\label{fig:dmeson2}
\end{figure}

In order to study  the dependence on the cutoff and coupling constants together with the isospin dependence of the in-medium DN interaction, on the l.h.s. of Fig.~\ref{fig:dmeson2} we show the  real and imaginary parts of the D-meson potential at $k_D=0$ as a function of the density.
In the case that only the D-meson is dressed (left panels), and fixing
$\Lambda=1$ GeV, the D-meson potential is governed by the $I=1$ component
while, when nucleons and pions are dressed (right panels), the $I=0$
component dominates due to the structure present at 2.8 GeV  in the
G-matrix. With regards to the dependence on the cutoff and coupling constant, we observe a weak dependence on the chosen set of parameters for both approaches. However, it is interesting to see that, for any chosen parameters, the coupled-channel effects seem to result in an overall  reduction of the in-medium effects independent of the in-medium properties compared to previous works \cite{qmc,arata,digal,amruta}. For example, when nucleons and pions are dressed in the self-consistent
process, the attraction lies in between $[2.6,-12.3]$ MeV for
$\Lambda=[0.8, 1.4]$ GeV at normal nuclear matter density $\rho_0$.

Finally, the spectral density at zero momentum is shown on the r.h.s. of Fig.~\ref{fig:dmeson2} for $\Lambda=1$ GeV and for several densities in the  two self-consistent approaches considered before. In the first approach (left panel), the quasiparticle peak moves slightly to  lower energies while the $\Lambda_c$ resonance is seen for energies around 1.63-1.65 GeV. For the second approach (right panel), the structure around 2.8 GeV of Fig.~\ref{fig:dmeson1} mixes with the quasiparticle peak. 

We have performed a microscopic self-consistent coupled-channel calculation of the D-meson spectral density embedded in symmetric nuclear matter assuming, as bare interaction, a separable potential \cite{tolos04}. We have obtained the $\Lambda(2593)_c$ dynamically and  we have concluded that the self-consistent coupled-channel effects result  in a small attractive real part of the in-medium potential for D-mesons. However, the production of D-mesons in the nuclear medium
will be still enhanced due to the broad D-meson
spectral density.  This effect is similar to the one obtained for the enhanced $\bar K$ production in heavy-ion collisions
\cite{Toloseffect}. The in-medium effects seen in this work can be studied with the future PANDA experiment at GSI \cite{ritman}.

\section*{Acknowledgment}

L.T. acknowledges financial support from the Alexander von Humboldt Foundation.

\end{document}